\documentclass[conference]{IEEEtran}
\IEEEoverridecommandlockouts
\usepackage{cite}
\usepackage{amsmath,amssymb,amsfonts}
\usepackage{algorithmic}
\usepackage{graphicx}
\usepackage{textcomp}
\usepackage{xcolor}
\usepackage{stfloats}
\usepackage{comment}
\def\BibTeX{{\rm B\kern-.05em{\sc i\kern-.025em b}\kern-.08em
    T\kern-.1667em\lower.7ex\hbox{E}\kern-.125emX}}
\begin{document}

\newcommand{\ipv}[1]{{\color{blue}[IPV:] #1}}

\title{Vertical Power Delivery for Emerging Packaging and Integration Platforms - Power Conversion and Distribution\\
\thanks{This work was supported by the Center for Heterogeneous Integration of Micro Electronic Systems (CHIMES), one of seven centers in Joint University Microelectronics Program (JUMP) 2.0, a Semiconductor Research Corporation (SRC) program sponsored by the Defense Advance Research Project Agency (DARPA).}}

\author{\IEEEauthorblockN{Sriharini Krishnakumar}
\IEEEauthorblockA{\textit{Electrical and Computer Engineering} \\
\textit{University of Illinois Chicago}\\
Chicago, IL, USA \\
skrish47@uic.edu}
\and
\IEEEauthorblockN{Inna Partin-Vaisband}
\IEEEauthorblockA{\textit{Electrical and Computer Engineering} \\
\textit{University of Illinois Chicago}\\
Chicago, IL, USA \\
vaisband@uic.edu}
}

\maketitle

\begin{abstract}
Efficient delivery of current from PCB to point-of-load (POL) is a primary concern in modern high-power high-density integrated systems. Traditionally, a 48~V power signal is converted to the low, POL voltage at the board and/or package level. As interconnect has become the dominant power loss component, minimizing voltage drop across the laterally routed portions of the board-to-die interconnect (referred to as horizontal interconnect) is a promising approach to enhance the efficiency of the power delivery system. Delivering lower current \emph{vertically}, at a higher voltage should therefore be considered. High-power conversion near POL, however, results in higher switching and inductor losses, exhibiting an undesired power efficiency tradeoff. To address this problem,  four vertical power delivery architectures are proposed in this paper, considering state-of-the-art power converter topologies, integration levels, and voltage conversion schemes. Embedding Silicon (Si) and Gallium Nitride (GaN) power devices and inductors on top of and/or within the interposer is investigated. Integrating GaN power devices on a dedicated power die is also discussed. Various multi-stage 48V-to-1V power conversion schemes are examined and state-of-the-art power conversion circuits are reviewed. Power delivery characteristics with these architectures are determined for a high power (1~kW) high-current density (2~A/mm\textsuperscript{2}) system.
\end{abstract}


\section{Introduction}
\begin{figure}[t!]
\centerline{\includegraphics[width=0.5\textwidth]{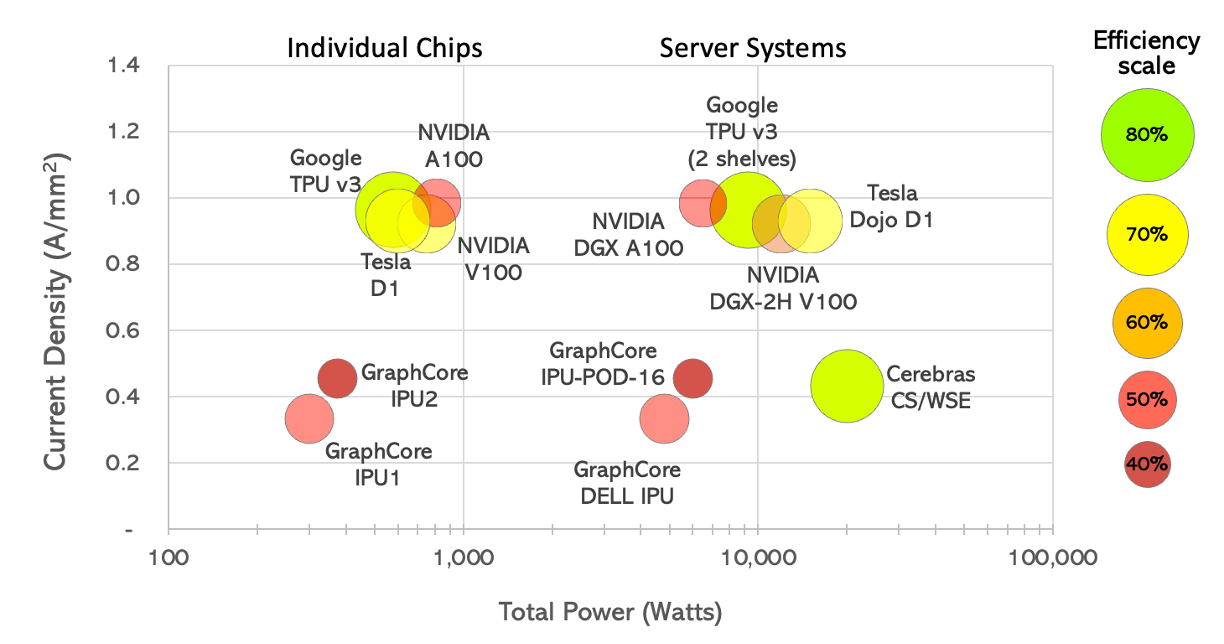}}
\caption{Power and current density demands in state-of-the-art HPC systems. The size (and color shade) of each data point corresponds to the efficiency of the power delivery system. Data for the individual chips and corresponding server systems are shown respectively on the left and right side of the figure.}
\label{fig:hpc_power}
\end{figure}

Modern three dimensional (3D) and heterogeneous 2.5D integration technologies and design methodologies have enabled powerful computing systems that operate at a low voltage ($<$1V). 
These high-performance integrated systems require high power to be delivered to points-of-load (POLs) at high current density. Latest power demand trends in high-performance computing (HPC) (e.g., for artificial intelligence (AI) applications) and data centers are shown in Figure \ref{fig:hpc_power}, confirming that the state-of-the-art AI accelerators are rapidly approaching a thousand watts for an individual chip and 20~kW for a server system \cite{b11}\cite{b12}\cite{b13}.
%
\begin{figure}[hbt]
\centerline{\includegraphics[width=0.5\textwidth]{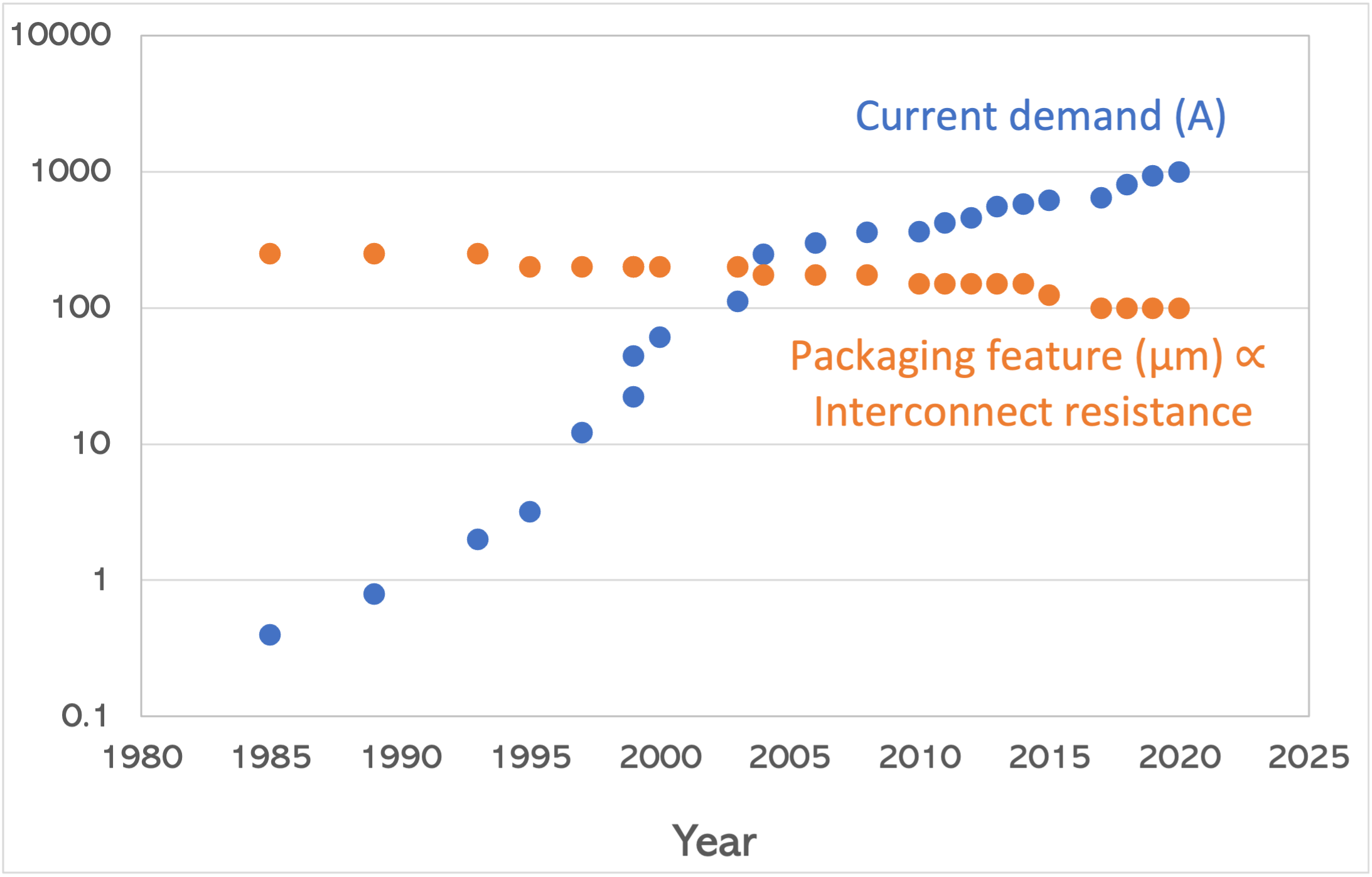}}
\caption{Current demand and PPDN resistance in high performance integrated systems. The current demand is estimated based on power density, as reported by Intel and a typical die size of 200 mm\textsuperscript{2}. The packaging feature data is taken from \cite{b10}.}
\label{fig:scaling}
\end{figure}

Traditionally, power efficient converters are utilized to convert high-voltage low-current power to POL-voltage high-current power at printed circuit board (PCB). State-of-the-art step-down switching mode power supplies (SMPSs) designed on PCB without stringent area and frequency constraints exhibit $>$90\% power efficiency \cite{b9}. The current at the output of a step-down SMPS converter increases linearly with the decrease in the output voltage. Thus, with PCB-level power conversion, high current is distributed \emph{laterally} through the multi-level packaging power distribution network (PPDN), from the converter output on the PCB, through the package, and to the POLs on the die. The dc loss in PPDN increases quadratically with the PPDN current, exhibiting high loss in high-performance applications. For example, a 48V-to-1V conversion at the PCB level, results in 48x higher PPDN current and, consequently, a 48\textsuperscript{2} times higher PPDN loss. For a typical PPDN resistance in the order of a few of milliohms, delivering thousand amperes from PCB to die would yield an impractical power loss of $I^2R > 4$ kW. Furthermore, power density in these modern HPC accelerators approaches 1 A/mm\textsuperscript{2}, as shown in Figure \ref{fig:hpc_power}, and is expected to double in the near future, imposing heavy area constraints on the components of the power delivery system. Existing packaging technology cannot support such current density and existing power conversion circuits, inductors, and capacitors cannot be designed within such a small form factor with high power efficiency. Thus, alternative power delivery architectures and power conversion topologies are required.

\begin{figure*}[b!]
\vspace{-10pt}
\centerline{\includegraphics[width=\textwidth]{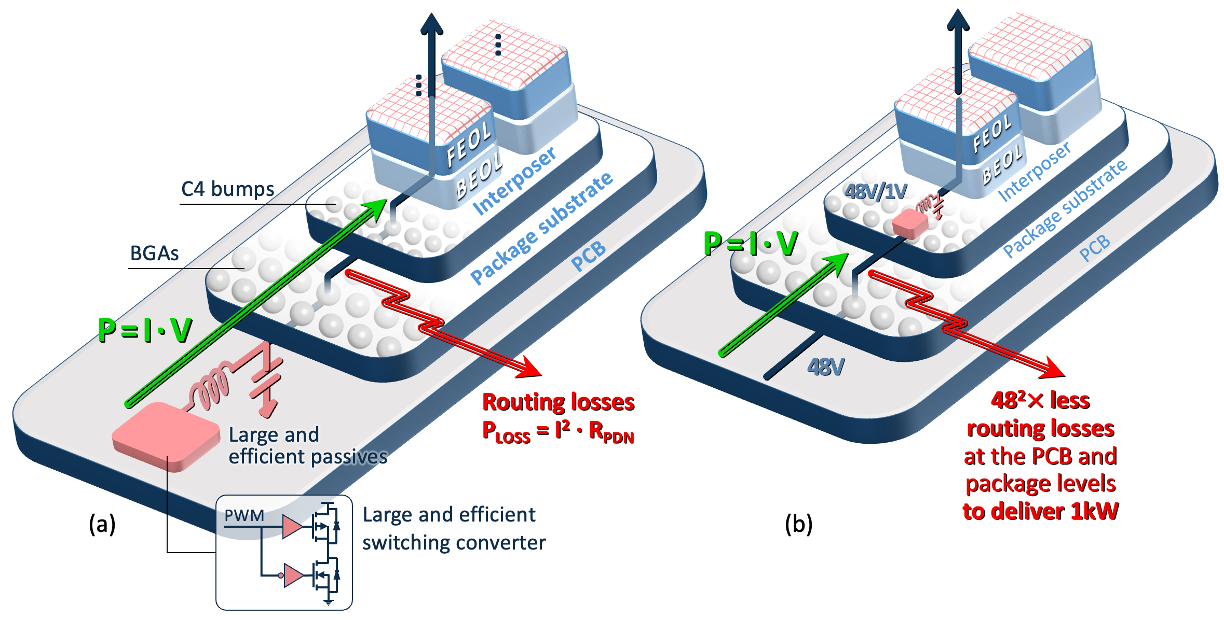}}
\caption{Illustration of power savings with voltage regulation on interposer as compared with the traditional PCB-level power conversion.}
\label{fig:vpd}
\end{figure*}

Naturally, reducing the overall PPDN resistance, $R_{PPDN}~=~\rho \cdot \frac{l}{A}$ ($\rho$, $l$, and $A$ are the interconnect resistivity, length, and cross-sectional area) should be considered. 
One approach to reduce the PPDN resistance is through increasing the interconnect density. 
Advanced vertical interconnect technologies (e.g., \textmu-bumps and Cu-Cu direct bonding \cite{b14}) have recently been demonstrated as a promising alternative to the traditional solder-based ball grid arrays (BGAs) and C4 bumps. However, these advances cannot mitigate the PPDN power loss which increases with $I^2$ --- while current demand has increased by several orders of magnitude over the last several decades, the packaging feature (which effectively determines the PPDN resistance) has only decreased by a factor of $\sim$4x, as shown in Figure \ref{fig:scaling}. Thus, the primary goal of these advanced packaging solutions is to increase the power density. 

Alternatively, significant power savings can be attained by shortening the long lateral distance between the output of the power converter and on-chip POLs. This approach is known as vertical power delivery where, (i) voltage regulators (VRs)  are integrated close to the POLs, and (ii) power distribution and conversion components are vertically aligned under the die(s). The concept of power savings with integrated voltage regulators is illustrated in Figure \ref{fig:vpd}.
While promising from power efficiency perspective, vertical power delivery poses fundamental novel challenges in power system design. To integrate high-ratio (e.g., 48V-to-1V) voltage regulators within a small form factor (in-package and/or in-interposer) while maintaining a high conversion efficiency, new circuit topologies co-designed with wide-band materials (e.g., GaN power transistors) are required. Furthermore, multi-stage voltage regulators distributed within and across different packaging levels should be considered. Finally, to mitigate the high design complexity of such multi-stage distributed power delivery system, accurate system-level models, power delivery architectures, and design methodologies are required.

\begin{table*}[t!]
  \caption{Typical characteristics of vertical interconnect}
  \begin{center}
    \begin{tabular}{l|r|l|l|r|r|r|r}
      \hline
      Packaging level & Platform area (mm\textsuperscript{2}) & Type & Material & Diameter (\textmu m) & Cross-area (\textmu m\textsuperscript{2}) & Height (\textmu m) & Pitch (\textmu m) \\
      \hline
      PCB/PKG & 1800 & BGAs & solder & 400 & 125,664 &300 & 800 \\
      PKG/Interposer & 1200 & C4 bumps & solder & 100 & 7,854 & 70 & 200 \\
      Through-Interposer & 1200 & TSVs & Cu & 5 & 20 & 50 & 10 \\
      Interposer/Die & 500 & \textmu-bumps & solder & 30 & 707 & 25 & 60 \\ 
      Interposer/Die & 500 & advance pad & Cu & -- & 100 & 10 & 20 \\
      \hline
    \end{tabular}
    \label{tab:vic_char}
  \end{center}
\end{table*}
\begin{figure}[hbt]
  \centerline{\includegraphics[width=0.5\textwidth]{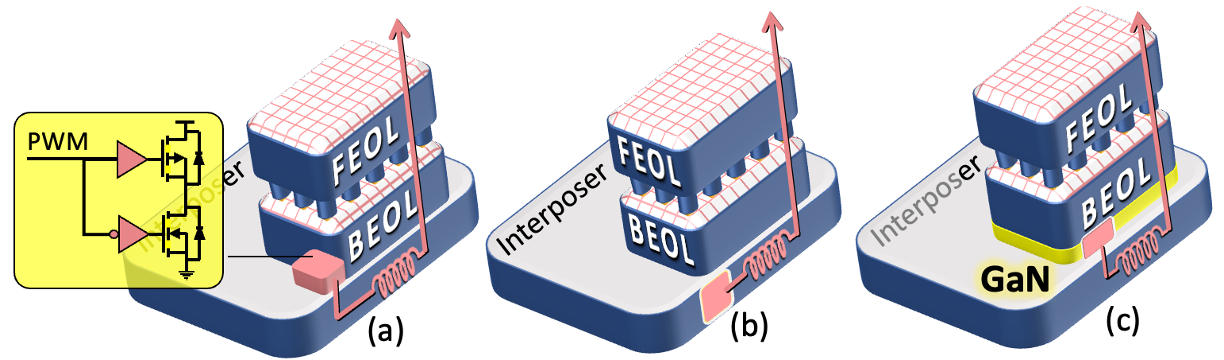}}
  \caption{Proposed vertical power delivery architectures, featuring (a) single-step high-voltage ratio power conversion with on-interposer power transistors and discrete capacitors and inductors embedded in-interposer, (b) a single-step high-voltage ratio power conversion with power transistors, capacitors, and inductors embedded in-interposer, (c) multi-stage power conversion comprising embedded passives, a high-voltage regulator with on-interposer power transistors, and a lower-voltage regulator with power transistors integrated below the functional die (e.g., on a dedicated stacked die). }
  \vspace{-10pt}
  \label{fig:vpd_arch_x3}
\end{figure}

\begin{figure*}[t!]

\centerline{\includegraphics[width=0.8\textwidth]{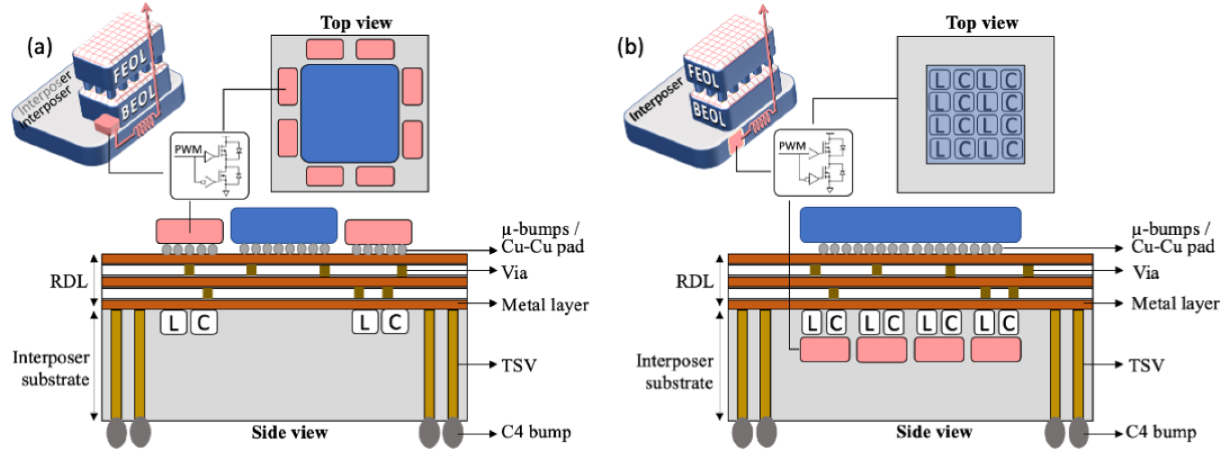}}
\caption{Distributed vertical power delivery with, (a) power transistors  distributed along the die periphery on interposer and discrete inductors and capacitors embedded in interposer, and (b) power transistors are embedded in interposer below the die and passives embedded in interposer, redistribution layer (RDL) below the die, or back-end-of-line (BEOL).}
\vspace{-10pt}
\label{fig:vpd_arch_distrib}
\end{figure*}

The rest of the paper is organized as follows. Four power delivery architectures are proposed in Section \ref{sec:vpd} for delivering power vertically in high-power (1 kW) high-current density (2 A/mm\textsuperscript{2}) systems.  Promising power conversion circuits are reviewed in Section \ref{sec:vrs}. Challenges and opportunities involved with GaN-based power conversion circuits are also discussed in this section. Characterization results of the proposed power delivery architectures are presented in Section \ref{sec:results}. The paper is concluded in Section \ref{sec:conclusions}.

\section{Vertical Power Delivery Architectures}\label{sec:vpd}

Delivering high-density high current ($>$1A/mm\textsuperscript{2}, $>$1kA) from PCB through multiple packaging levels exhibits a tremendous power loss (e.g., over 30\% power loss has recently been reported in state-of-the-art AI accelerators \cite{b11}) and is limited by insufficient density of vertical interconnect. A promising solution is to deliver high-voltage low-current power through packaging levels and convert at POL or near-POL. However, integration of VRs closer to POLs, yields undesired tradeoffs between power distribution and power conversion losses. A tradeoff-aware exploration of the power delivery architecture space is required to determine a preferred architecture for a given application. Several promising architectures are proposed and analyzed in this section.

The proposed power delivery architectures are shown in Figure \ref{fig:vpd_arch_x3}. The first two architectures ($A_1$ and $A_2$, as shown in Figures \ref{fig:vpd_arch_x3}(a) and \ref{fig:vpd_arch_x3}(b), respectively) introduce a single-stage 48V-to-1V power conversion with on-/in-interposer VR. The third architecture introduces a multi-stage power conversion. In this work, the third architecture is considered with embedded in-interposer passives and on-interposer power transistors for high voltage regulation. The lower-voltage regulation is achieved with power transistors integrated even closer to POL (e.g., on a dedicated power die beneath the functional die \cite{b8}). Both 48V-to-12V and 48V-to-6V and, consequently, the 12V-to-1V and 6V-to-1V configurations are evaluated for, respectively, the higher- and lower-voltage regulation. These architectures are referred to as $A_{3@12V}$ and $A_{3@6V}$. In the reference architecture ($A_0$), 48V-to-1V conversion is considered at the PCB level. 

Note that to efficiently deliver a thousand watts with 1~A/mm\textsuperscript{2} current density, multiple VRs should be distributed within the individual packaging levels and all the components of the power delivery system should be maximally vertically aligned. Preferred distribution scheme of the individual VR components varies at different packaging levels. In this work, the on-interposer power transistors in architectures $A_1$ and $A_3$ are distributed uniformly along the periphery of the die. The passives in this architecture are also placed along the perimeter of the die, below the power transistors, embedded in the interposer. Alternatively, the in-interposer and in-power die power transistors in architectures $A_2$ and $A_3$ respectively are uniformly distributed below the die along with the corresponding passives embedded in-interposer, occupying $\sim$50\% of the die area in the interposer. The concept of distributed vertically aligned power delivery is illustrated in Figure \ref{fig:vpd_arch_distrib}.

Vertical interconnect characteristics for evaluating the proposed power delivery architectures are determined based on \cite{b8}, as listed in Table \ref{tab:vic_char}. The number of vertical interconnect components at various packaging levels is determined based on these characteristics and die size of 1~kA~/~2~A/mm\textsuperscript{2}~=~500~mm\textsuperscript{2}. Both power and ground distribution networks are considered.

\section{Compact Power-Efficient High-Ratio Voltage Converters}\label{sec:vrs}
\begin{figure*}[b!]
\centerline{\includegraphics[width=\textwidth]{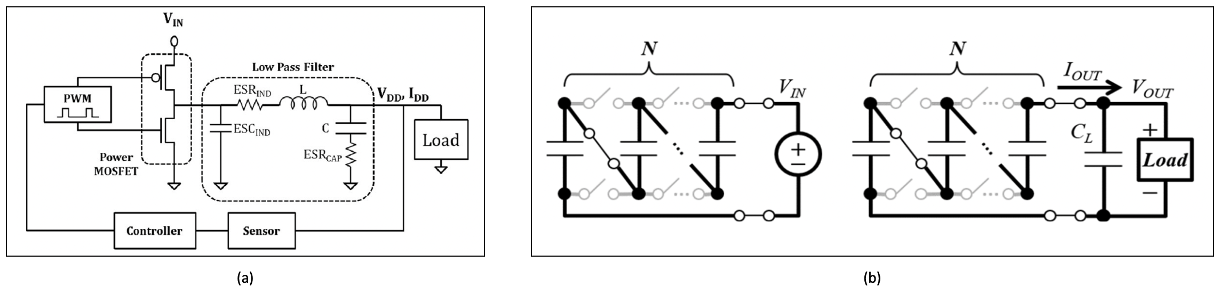}}
\caption{Power converters, (a) SMPS buck converter circuit \cite{b16}, and (b) SC series-parallel charge pump with flying capacitors connected in series to the input during the phase 1 (on the left) in parallel to the load during phase 2 (on the right)\cite{b17}}.
\label{fig:SMPS_SC}
\end{figure*}

Historically transformer-based converters were utilized at the PCB and/or package level to efficiently bring down the voltage, while operating at lower frequencies for reduced switching loss. However, transformer-based converters are bulky and thus impractical for integration within a small form factor in vertical power delivery systems. Thus, alternative power converter architectures need to be considered. A basic topology of a SMPS and a switched-capacitor (SC) converters is shown in Figure \ref{fig:SMPS_SC}. 

Power loss in a typical SMPS converter comprises switching power loss due to periodic switching of the power transistors, conduction power loss dissipated in power transistors due to their non-zero effective drain-to-source resistance, and inductor and capacitor power loss. Integrated passives limited by the small form factor exhibit lower energy capacity and need to be switched faster.  Switching loss increases linearly with switching frequency, thus posing a significant challenge in efficiently converting the power at POL. Alternatively, high-ratio voltage conversion, such as 48V-to-1V requires a pulse width modulator (PWM) with an ultra-low on-time ($\sim$2\%), limiting the switching frequency of the converter. Thus, while buck converter exhibits high efficiency in delivering high current and regulating the output voltage, its efficiency is limited to medium-to-low voltage conversion ratio. 

Alternatively, while power losses of a SC converter at high frequencies are also dominated by switching loss, a low on-time is not required in high-ratio SC topologies. Thus, SC converters should be preferred over SMPS topologies for stepping down high voltages. A comparative study between various buck-derived and SC-derived topologies is presented in this section based on recent state-of-the-art converters \cite{b2}\cite{b3}\cite{b4}. A primary goal is to understand which converters maximize the power efficiency within the system-level constraints of the individual power delivery architectures. The following voltage conversion topologies are considered in this work as part of the power delivery architectures, as described in Section \ref{sec:vpd}.


\emph{Double series capacitor hybrid (DSCH) converter} is a buck derived VR. With this (DSCH) converter a high input voltage is efficiently stepped down to 1/3 of the input voltage with a compact SC, comprising two capacitors and a switch \cite{b5}. Consequently, the lower voltage is converted to a dual-phase POL voltage with a buck-based VR. Similar to a dual phase synchronous buck converter, this topology comprises a small number of switches and passive components. At the same time, the challenge of high switching losses due to high input voltage and ultra-low on-time is efficiently mitigated with this SC-like sub-circuit. A primary challenge is the difference in current between the two conversion phases, resulting in higher conduction loss. DSCH converters occupy less area and are more suitable for lower conversion ratios such as 12V-to-1V or 6V-to-1V and are preferred for the second stage conversion in a multi-stage power delivery architecture, as shown in Figure \ref{fig:vpd_arch_x3}(c). Alternatively, for high-ratio 48V-to-1V conversion, a maximum load current of 30 A and power efficiency of 91.5\% at 10 A are reported using Si power transistors in \cite{b5}. To increase power density and efficiency of the converter at high frequency, GaN power transistors (which are known for their high electron mobility) are considered in this paper. 

\emph{Dual phase multi-inductor hybrid (DPMIH) converter } is a SC based VR that comprises eight switches, four inductors, and three capacitors \cite{b6}, overcoming two primary limitations of traditional SC VRs: 1) hard switching between two capacitors (or a capacitor and a voltage source) charged to different voltage levels, and 2) discrete conversion ratio. With the DPMIH topology, soft switching is enabled by connecting each of the capacitors to an inductor, balancing different voltage levels across capacitors without the need for an external control. For a 48V-to-1V conversion, a maximum load current of 100 A and peak efficiency of 90.9\% at 30 A has been reported with GaN power transistors in \cite{b6}. However, the topology exhibits higher area overhead due to larger number of required inductors. Thus, DPMIH VRs should be preferred for converting a higher input voltage and delivering higher output current, within a larger form factor. In this paper, DPMIH VRs are considered for a single-stage 48V-to-1V conversion with architectures $A_1$ and $A_2$, as well as for the first-stage (48V-to-12V or 48V-to-6V) conversion in the multi-stage architectures $A_{3@12V}$ and $A_{3@6V}$. 

\emph{Three-level hybrid Dickson (3LHD) converter} is a three-phase VR compring eleven switches, five flying capacitors, and three inductors to effectively reduce conduction loss \cite{b7}. All the flying capacitors are self-balanced, and do not require additional control, thus reducing the complexity in the control circuitry. In addition, the input voltage is stepped down by 10x (e.g., to 4.8 V) with this approach, effectively reducing power transistor stress and switching loss while increasing the on-time from 2\% to 20\%. For the 48V-to-1V conversion, a maximum load current of 12 A and peak efficiency of 90.4\% at 3A is reported with two GaN and nine Si power transistors in \cite{b7}. This topology is evaluated in this paper with all eleven GaN switches integrated at various levels of packaging hierarchy. Note that while eleven switches are used in this topology, the area occupied by all the switches is lower when compared to DPMIH converter topology due to lower volume of the delivered current.
\begin{figure*}[b!]
\centerline{\includegraphics[width=\textwidth]{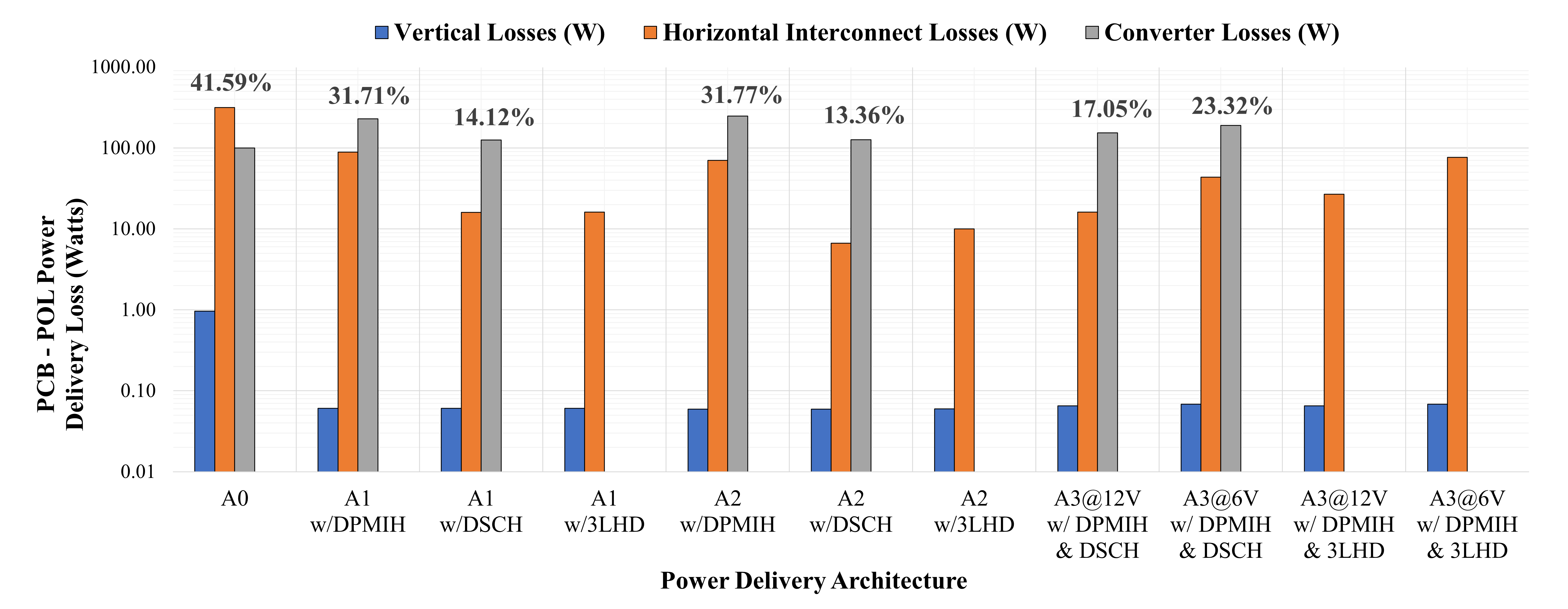}}
\caption{PCB-to-POL power loss with the proposed power delivery architectures. The power is assumed to be delivered to PCB at 48 V. The die operates at 1 V, 1 kA. The total power delivery loss is shown as per cent of the total power (1 kW) available at the PCB.}
\label{fig:power_loss}
\end{figure*}

\section{Results and Discussion}\label{sec:results}
Characteristics of the DSCH, DPMIH, and 3LHD VR topologies, as described in Section \ref{sec:vrs}  with all GaN power transistors are listed in Table \ref{tab:VRs} with respect to the power delivery architectures, as described in Section \ref{sec:vpd}. Note, that the passive components required for the power conversion are assumed to fit in the area occupied by their switches, as illustrated in Figure \ref{fig:vpd_arch_distrib}(b). Also note, that in practice, the passive components can be (partially) integrated within the redistribution layers (RDL). While embedded state-of-the-art inductors only support current density of up to 1 A/mm\textsuperscript{2} \cite{b15}, the performance of capacitors and inductors is expected to sufficiently improve in the near-term future. 

\begin{table}[t!]
  \caption{Characteristics of state-of-the-art compact high-current 48V-to-1V power converters.}
  \begin{center}
    \begin{tabular}{l|c|c|c}
      \hline
      & DPMIH & DSCH & 3LHD \\
      \hline
      Conversion scheme & 48V-to-1V & 48V-to-1V & 48V-to-1V \\
      Max load current & 100 A & 30 A & 12 A \\
      Peak efficiency & 90.0\% & 91.5\% & 90.4\% \\
      Current at peak efficiency & 30 A & 10 A & 3 A \\
      Number of switches & 8 & 5 & 11 \\
      Number of switches per mm\textsuperscript{2} & 0.15 & 0.69 & 1.22 \\
      Number of inductors & 4 & 2 & 3 \\
      Total inductance & 4 \textmu H & 0.88 \textmu H & 1.86 \textmu H \\
      Number of capacitors & 3 & 2 & 5 \\
      Total capacitance & 15 \textmu F & 6.6 \textmu F & 5 \textmu F \\
      VRs along die periphery & 8 & 48 & 48 \\
      VRs below the die  & 7 & 48 & 48 \\
      \hline
    \end{tabular}
    \label{tab:VRs}
  \end{center}
\end{table}

The four proposed power delivery architectures ($A_1$, $A_2$, $A_{3@12V}$, and $A_{3@6V}$) are modeled and dc power loss is determined with the DPMIH, DSCH, and 3LHD VR topologies. The reference architecture ($A_0$) is modeled with a 90\% efficient 48V-to-1V converter, comprising a transformer-based 48V-to-12V first-stage converter and a multi-phase synchronous 12V-to-1V second-stage buck converter. In the other architectures, power transistors are distributed along the periphery and/or below the die.
Note that based on the maximum output current and area, as determined by the individual VR topologies (see Table \ref{tab:VRs}), a different number of converters are required.
When the required number of converters cannot be fitted along the periphery of the die additional rows of VRs are utilized farther away from the perimeter of the die.

The distribution of power loss among vertical interconnect (BGAs, C4 bumps, µ-bumps), horizontal interconnect, and VR components is shown in Figure \ref{fig:power_loss}. The total dissipated power from  PCB-to-POL is shown on the plot as per cent of the total power (1 kW) available at the PCB. Since the efficiency for the required current load of ~20 A per VR is not reported in \cite{b7}, power loss for the proposed architectures with the 3LHD topology is not shown in Figure \ref{fig:power_loss}.
Based on the results, the overall power loss is dominated by the VRs and horizontal interconnect while power dissipated in vertical interconnect is negligible. As expected, significant power savings are possible with vertical power delivery --- while the traditional approach exhibits over 40\% power loss, most of the proposed architectures exhibit promising efficiency of $\sim$80\%. Furthermore, power density is also significantly improved with vertical power delivery. For example, assuming that vertical power interconnect cannot exceed 60\% and 85\% of all the available BGAs and C4 bumps, respectively, an unreasonably large die of 1,200 mm\textsuperscript{2} is required with the reference architecture to distribute 1 kA current through vertical interconnect, limiting the power density of $A_0$ to 0.8 A/mm\textsuperscript{2}. Alternatively, with vertical approach, 1 kA current can be delivered to a 500 mm\textsuperscript{2} die (i.e., with 2 A/mm\textsuperscript{2} density), while utilizing only 1\% of BGAs, 2\% of C4 bumps, 10\% of TSVs, and $<$20\% of advanced Cu-Cu pads.

While horizontal power loss is reduced by up to 19x and 7x with the architectures $A_{3@12V}$ and $A_{3@6V}$ respectively, the dual-stage power conversion yields a lower power efficiency when compared to the single-stage conversion approach in architectures $A_1$ and $A_2$ with DSCH and 3LHD converters. Note that albeit the similarities between the architectures $A_1$ and $A_2$ with DSCH or 3LHD converters, the distribution of current load among the converters significantly differ between the two architectures. With $A_1$, the current delivered by various converters varies between 16 and 27 amperes. Alternatively, with $A_2$, the individual converters placed below the center of the die provide as much as 93 amperes per VR while others provide as little as 10 amperes per VR. A much broader range of current load therefore has to be supported by the converters in $A_2$.



\section{Conclusions}\label{sec:conclusions}
Delivering high-power vertically and converting power closer to POLs is a promising approach for increasing the overall efficiency of the power delivery system above 80\%. However, integrating high-voltage converters within a small form factor is challenging and exhibits low power efficiency when compared to the voltage regulators implemented at the PCB. Four novel vertical power delivery architectures are proposed and characterized for high power (1~kW) high-current density (2 A/mm\textsuperscript{2}) systems. Three state-of-the-art voltage regulators are reviewed and compared. All the proposed architectures, as considered with the state-of-the-art VRs, exhibit power loss of $<$10\% in PPDN and $>$10\% in the converters. It is vital to improve the efficiency of the converters delivering higher currents, to improve the efficiency of the next generation high-performance systems.


\end{document}